\documentclass[11pt]{article}
\usepackage{graphicx}
\usepackage{amssymb}
\usepackage{graphics}

\input{tcilatex}
\begin{document}

\title{$V_{\mathrm{ud}}$ from neutron beta decay}
\author{Oliver Zimmer \\
Institut Laue Langevin, 38042 Grenoble, France}
\maketitle

\begin{abstract}
\noindent The experimental determination of $V_{\mathrm{ud}}$ from neutron
beta decay requires accurate values of the neutron lifetime and the ratio of
the weak axial-vector to vector coupling constants of the nucleon. The
latter is derived from measurements of angular correlation coefficients in
the differential decay probability, such as the beta asymmetry parameter $A$%
, the neutrino-electron angular correlation coefficient $a$, or the
parameter $C$ of the proton asymmetry. As a probe free from nuclear
structure corrections the decay of the free neutron has the potential to
provide the most accurate value of $V_{\mathrm{ud}}$. Towards that end
however, the experimental sensitivity still needs to be further improved to
become competitive with superallowed nuclear beta decays. This contribution
briefly reviews the current status of those neutron decay studies relevant
for the determination of $V_{\mathrm{ud}}$.\bigskip

\noindent Proceedings of CKM 2012, the 7th International Workshop on the CKM
Unitarity Triangle, University of Cincinnati, USA, 28 September - 2 October
2012\bigskip
\end{abstract}

\noindent The semi-leptonic beta decay of the free neutron is a mixed
transition with exactly known Fermi and Gamow-Teller matrix elements given
by Clebsch Gordan coefficients \cite{Dubbers/2012,Abele/2008}, and with
negligible isospin symmetry breaking effects in the neutron-to-proton vector
current transition matrix element \cite{Kaiser/2001}. As from superallowed $%
0^{+}\rightarrow 0^{+}$ nuclear beta decays \cite{Melconian/2012}, nuclear
mirror transitions \cite{Naviliat/2012} and pion beta decay \cite%
{Pocanic/2004}, $V_{\mathrm{ud}}$ is determined from the weak vector
coupling constant of the nucleon, $g_{\mathrm{V}}=V_{\mathrm{ud}}G_{\mathrm{F%
}}$, with the Fermi coupling constant $G_{\mathrm{F}}$ determined from the
muon lifetime \cite{Gorringe/2012}. In the standard $V-A$ theoretical
description of neutron decay the axial-vector current contributes with an
empirical coupling constant $g_{\mathrm{A}}$, so that two observables are
needed to access $V_{\mathrm{ud}}$. These are the neutron lifetime,%
\begin{equation}
\tau _{\mathrm{n}}^{-1}\propto g_{\mathrm{V}}^{2}+3g_{\mathrm{A}}^{2},
\end{equation}%
and a coefficient appearing in the differential decay probability expressed
in terms of various angular correlations \cite{Jackson/1957}. Highest
experimental sensitivity has been achieved for the beta asymmetry
coefficient $A$, which within the standard model with negligible time
reversal invariance violation is given by%
\begin{equation}
A=-2\frac{\lambda +\lambda ^{2}}{1+3\lambda ^{2}},
\end{equation}%
where%
\begin{equation}
\lambda =\frac{g_{\mathrm{A}}}{g_{\mathrm{V}}}.
\end{equation}%
$V_{\mathrm{ud}}$ is derived from the following expression given by Marciano
and Sirlin, 
\begin{equation}
\left\vert V_{\mathrm{ud}}\right\vert ^{2}=\frac{4908.7\left( 1.9\right) \;%
\mathrm{s}}{\tau _{\mathrm{n}}\left( 1+3\lambda ^{2}\right) },  \label{MS}
\end{equation}%
where the numerical value quoted in parentheses represents the uncertainty
estimate of their calculation of the radiative correction $\Delta _{R}$
accounting for short-distance loop effects \cite{Marciano/2006}. The
theoretical knowledge of $\Delta _{R}$ ultimately limits the accuracy to
currently $\delta \left\vert V_{\mathrm{ud}}\right\vert =1.9\times 10^{-4}$.
While the experimental precision of superallowed $0^{+}\rightarrow 0^{+}$
nuclear beta decay studies has been pushed beyond this limitation \cite%
{Melconian/2012,Hardy/2009,Towner/2010}, in neutron decay studies there is
still large room for improvement. It is reasonable to request that the
contributions from $\tau _{\mathrm{n}}$ and $\lambda $ to the total error of 
$V_{\mathrm{ud}}$ should be made smaller than that due to $\Delta _{R}$.
This defines the goal that $\tau _{\mathrm{n}}$ and $\lambda $ need to be
measured with accuracies of at least $\delta \tau _{\mathrm{n}}\leq 0.34$ s
and $\delta \lambda \leq 3\times 10^{-4}$.

The experimental situation of the neutron lifetime has been particularly
unsatisfactory. Until 2005, the world average of measured values looked
fine, $885.7\pm 0.8$ s with a reduced $\chi ^{2}=0.76$ \cite{Severijns/2006}%
. In 2005, Serebrov et al. published a new result of $878.5\pm 0.8$ s,
disagreeing by $6.5$ standard deviations \cite{Serebrov/2005,Serebrov-1/2008}%
. The Particle Data Group (PDG) \cite{Amsler/2008} said: \textquotedblleft
The most recent result, that of Serebrov 05, is so far from other results
that it makes no sense to include it in the average. It is up to workers in
this field to resolve this issue\textquotedblright . In the meantime, a new
result of $880.7\pm 1.8$ s became published \cite{Pichlmaier/2010}, and also
a still unpublished result of a magnetic trapping experiment \cite%
{Ezhov/2009} seems to support a lower lifetime value than the one accepted
before 2005. A closer look on the situation before 2005 shows that the world
average value was dominated by a single experiment with a quoted result of $%
885.4\pm 0.9_{\mathrm{stat}}\pm 0.4_{\mathrm{syst}}$ s \cite{Arzumanov/2000}%
. After publication of a Monte Carlo analysis of that experiment \cite%
{Serebrov-2/2008}, the authors scrutinized their procedures and recently
published a corrected, lower neutron lifetime value of $881.6\pm 0.8_{%
\mathrm{stat}}\pm 1.9_{\mathrm{syst}}$ s \cite{Arzumanov/2012}. With these
changes included, the most recent PDG average value is now $880.1\pm 1.1$ s 
\cite{Beringer/2012}. However, a scale factor of $S=1.8$ applied by the PDG
indicates that systematic uncertainties were not properly taken into account
in all experiments.

The experiments with strongest impact on the world average value were
performed using storage of ultracold neutrons (UCN) in material bottles.
There, the main challenge is to cope in a reliable manner with UCN losses
during wall collisions. A generally adopted strategy employs a variation of
trap size and/or mean UCN velocity in order to repeat the trapping
experiment with different mean times $t_{\mathrm{f}}$ of free flight between
wall collisions. The neutron lifetime value is obtained from an
extrapolation of measured UCN storage time constants to $t_{\mathrm{f}%
}^{-1}\rightarrow 0$. While several experiments involved extrapolations by
more than $100$ seconds, in the experiment \cite{Serebrov/2005} the closest
data point was off by only $\Delta t\approx 5$ s, thanks to small UCN losses
of vessel walls coated with a fluorinated oil, liquid at low-temperature.
Storage time constants almost as good were previously measured by the same
group in a lifetime experiment using solid oxygen as wall material \cite%
{Nesvizhevsky/1992}. However, the fact that its result of $888.4\pm 3.1$ s
is more than three standard deviations off the new result illustrates that,
despite small corrections, at least in one of these experiments errors were
estimated too optimistically. Also the result $886.3\pm 1.2_{\mathrm{stat}%
}\pm 3.2_{\mathrm{syst}}$ s \cite{Nico/2005} of the only still competitive
neutron lifetime experiment using a cold beam instead of UCN trapping is in
disagreement but will soon take data with an improved method to determine
the fluence of the beam. Magnetic trapping offers a route to determine the
neutron lifetime free from uncertainties associated with interactions of UCN
with material walls. Experiments well under way are \cite%
{Ezhov/2009,Huffman/2000,Shaughnessy/2009}, and several new projects were
begun around the world \cite%
{Walstrom/2009,Leung/2009,Zimmer-4/2000,Materne/2009}.

Like the neutron lifetime, also the neutron beta asymmetry was measured many
times by several independent groups (see, e.g. the reviews \cite%
{Dubbers/2012,Abele/2008}). While in earlier measurements corrections on the
raw asymmetry in the order of $15-30\%$ had to be applied for neutron
polarisation, magnetic mirror effects, solid angle and backgrounds,
experimental techniques have strongly been refined in the meantime, so that
in latest measurements of the PERKEO collaboration instrumental corrections
stayed at the level of experimental uncertainties. As an example of the
progress made, the last experiment on $A$ with the spectrometer PERKEO II
was performed with a neutron beam polarisation of $99.7(1)\%$ \cite%
{Mund/2012}, using now well established techniques of neutron polarisation
and polarimetry \cite{Klauser/2012,Kreuz/2005,Zimmer-1/1999,Zimmer-2/1999}.

The latest value for $\lambda $ compiled by the PDG is $\lambda =-1.2701\pm
0.0025$ ($S=1.9$) \cite{Beringer/2012}. Not yet included is a new result
from PERKEO II \cite{Mund/2012}, $\lambda =-1.2755\pm 0.0013$. The
determination of $\lambda =-1.27590_{-0.00445}^{+0.00409}$ with UCNA \cite%
{Liu/2010}, the only experiment on $A$ performed with ultracold neutrons and
described in impressive detail in \cite{Plaster/2012}, is in agreement with
this result, and work to further reduce uncertainties is ongoing. An
accuracy of $\delta \lambda =\pm 0.00067$ has been announced for a recent
new measurement of $A$ with the new PERKEO III spectrometer \cite%
{Maerkisch/2009,Maerkisch/2012} for which a blind analysis is underway.

Complementary to the beta asymmetry $A$, neutron decay offers further
observables sensitive to the ratio $\lambda $. Within the standard $V-A$
theory the neutrino-electron angular correlation coefficient $a$ is given by%
\[
a=\frac{1-\lambda ^{2}}{1+3\lambda ^{2}}. 
\]%
Two prior measurements of $a$ \cite{Stratowa/1978,Byrne/2002} contribute
only weakly to the global PDG value for $\lambda $ but new experiments are
currently taking data, using the spectrometers $a$SPECT \cite%
{Zimmer-1/2000,Glueck/2005,Simson/2009,Konrad/2009} and $a$CORN \cite%
{Wietfeldt/2009}. The proton asymmetry, for which recoil order, Coulomb and
model-independent order-$\alpha $ radiative corrections were calculated by
Glueck \cite{Glueck/1996}, depends on $\lambda $ as%
\[
C=0.27484\frac{4\lambda }{1+3\lambda ^{2}}. 
\]%
It has so far been measured only a single time \cite{Schumann/2008} and will
be measured soon again with PERKEO III and $a$SPECT and is also in the focus
of the PANDA proposal \cite{Chupp/2008}.

Marciano and Sirlin stated in 2006: "Future precision measurements of $\tau
_{\mathrm{n}}$ and $\lambda $ (...) will ultimately be the best way to
determine $V_{\mathrm{ud}}$, but for now it is not competitive". Today this
statement is still valid but there is visible progress and several ambitious
projects are in the pipeline. The present generation of existing neutron
decay spectrometers should lead to accuracies close to the goal set by
present theory limitations, and projected new spectrometers such as PERC 
\cite{Dubbers/2008} and $Nab$ \cite{Pocanic/2009} have the potential to go
well beyond.


\begin{thebibliography}{99}
\bibitem{Dubbers/2012} D. Dubbers and M.G. Schmidt, Rev. Mod. Phys. \textbf{%
83} (2011) 1111.

\bibitem{Abele/2008} H. Abele, Prog. Nucl. Phys. \textbf{60} (2008) 1.

\bibitem{Kaiser/2001} N. Kaiser, Phys. Rev. C \textbf{64} (2001) 028201.

\bibitem{Melconian/2012} D. Melconian, Proceedings to this workshop.

\bibitem{Naviliat/2012} O. Naviliat-Cuncic, Proceedings to this workshop.

\bibitem{Pocanic/2004} D. Pocanic et al., Phys. Rev. Lett. \textbf{93}
(2004) 181803.

\bibitem{Gorringe/2012} T. Gorringe, Proceedings to this workshop.

\bibitem{Jackson/1957} J.D. Jackson, S.B. Treiman, H.W. Wyld, Phys. Rev. 
\textbf{106} (1957) 517.

\bibitem{Marciano/2006} W.J. Marciano and A. Sirlin, Phys. Rev. Lett. 
\textbf{96} (2006) 032002.

\bibitem{Hardy/2009} J.C. Hardy and I.S. Towner, Phys. Rev. C \textbf{79}
(2009) 055502.

\bibitem{Towner/2010} I.S. Towner and J.C. Hardy, Rep. Prog. Phys. \textbf{73%
} (2010) 046301.

\bibitem{Severijns/2006} N. Severijns, M. Beck, O. Naviliat-Cuncic, Rev.
Mod. Phys. \textbf{78} (2006) 991.

\bibitem{Serebrov/2005} A. Serebrov et al., Phys Lett. B \textbf{605} (2005)
72.

\bibitem{Serebrov-1/2008} A.P. Serebrov et al., Phys. Rev. C \textbf{78}
(2008) 035505.

\bibitem{Amsler/2008} C. Amsler et al. (Particle Data Group), Phys. Lett. B 
\textbf{667} (2008) 1.

\bibitem{Pichlmaier/2010} A. Pichlmaier, V. Varlamov, K. Schreckenbach, P.
Geltenbort, Phys. Lett. B \textbf{693} (2010) 221.

\bibitem{Ezhov/2009} V.F. Ezhov et al., Nucl. Instr. Meth. A \textbf{611}
(2009) 167.

\bibitem{Arzumanov/2000} S. Arzumanov et al., Phys. Lett. B \textbf{483}
(2000) 15.

\bibitem{Serebrov-2/2008} A.P. Serebrov, A.K. Fomin, arXiv:0808.3975 (2008).

\bibitem{Arzumanov/2012} S.S. Arzumanov et al., Analysis and correction of
the measurement of the neutron lifetime, Pis'ma v Zh. E. T. Fiz. \textbf{95}
(2012) 248, JETP Lett. \textbf{95} (2012) 224.

\bibitem{Beringer/2012} J. Beringer et al. (Particle Data Group), Phys. Rev.
D \textbf{86} (2012) 010001.

\bibitem{Nesvizhevsky/1992} V. Nesvizhevsky, A. Serebrov et al., JETP 
\textbf{75} (1992) 405.

\bibitem{Nico/2005} J.S. Nico et al., Phys. Rev. C \textbf{71} (2005) 055502.

\bibitem{Huffman/2000} P.R. Huffman, C.R. Brome, J.S. Butterworth, K.J.
Coakley et al., Nature \textbf{403} (2000) 62.

\bibitem{Shaughnessy/2009} C.M. Shaughnessy et al., Nucl. Instr. Meth. A 
\textbf{611} (2009) 171.

\bibitem{Walstrom/2009} P.L. Walstrom et al., Nucl. Instr. Meth. A \textbf{%
599} (2009) 82.

\bibitem{Leung/2009} K.K.H. Leung and O. Zimmer, Nucl. Instr.Meth. A \textbf{%
611} (2009) 181.

\bibitem{Zimmer-4/2000} O. Zimmer, J. Phys. G: Nucl. Part. Phys. \textbf{26}
(2000) 67.

\bibitem{Materne/2009} S. Materne et al., Nucl. Instr. Meth. A \textbf{611}
(2009) 176.

\bibitem{Mund/2012} D. Mund et al., arXiv:1204.0013 (2012).

\bibitem{Klauser/2012} C. Klauser et al., Phys. Conf. Ser. \textbf{340}
(2012) 012011.

\bibitem{Kreuz/2005} M. Kreuz et al., Nucl. Instr. Meth. A \textbf{547}
(2005) 583.

\bibitem{Zimmer-1/1999} O. Zimmer, Phys. Lett. B \textbf{461} (1999) 307.

\bibitem{Zimmer-2/1999} O. Zimmer et al., Phys. Lett. B \textbf{455} (1999)
62.

\bibitem{Liu/2010} J. Liu et al., Phys. Rev. Lett. \textbf{105} (2010)
181803.

\bibitem{Plaster/2012} B. Plaster et al., arXiv:1207.5887 (2012).

\bibitem{Maerkisch/2009} B. Maerkisch et al., Nucl. Instr. Meth. A \textbf{%
611 }(2009) 216.

\bibitem{Maerkisch/2012} B. Maerkisch, private communication.

\bibitem{Stratowa/1978} C. Stratowa et al., Phys. Rev. D \textbf{18} (1978)
3910.

\bibitem{Byrne/2002} J. Byrne et al., J. Phys. G: Nucl. Part. Phys. \textbf{%
28} (2002) 1325.

\bibitem{Zimmer-1/2000} O. Zimmer et al., Nucl. Instr. Meth. A \textbf{440}
(2000) 548.

\bibitem{Glueck/2005} F. Glueck et al., Eur. Phys. J. A \textbf{23} (2005)
135.

\bibitem{Simson/2009} M. Simson et al., Nucl. Instr. Meth. A \textbf{611}
(2009) 203.

\bibitem{Konrad/2009} G. Konrad et al., Nucl. Phys. A \textbf{827 }(2009)
529.

\bibitem{Wietfeldt/2009} F.E. Wietfeldt et al., Nucl. Instr. Meth. A \textbf{%
611} (2009) 207.

\bibitem{Glueck/1996} F. Glueck, Phys. Lett. B \textbf{376} (1996) 25.

\bibitem{Schumann/2008} M. Schumann et al., Phys. Rev. Lett. \textbf{100}
(2008) 151801.

\bibitem{Chupp/2008} T. Chupp et al., PANDA@SNS proposal,
http://research.physics.lsa.umich.edu/chupp/.

\bibitem{Dubbers/2008} D. Dubbers et al., Nucl. Instr. Meth. A \textbf{596}
(2008) 238.

\bibitem{Pocanic/2009} D. Pocanic et al., Nucl. Instr.Meth. A \textbf{611}
(2009) 211.
\end{thebibliography}
\end{document}